\documentclass[aps,prd,epsf,showpacs,amsmath,amssymb,graphics,
twocolumn,10pt]{revtex4}
\usepackage{graphicx}
\usepackage{dcolumn}
\usepackage{bm}

\newcommand{\be}{\begin{equation}}
\newcommand{\ee}{\end{equation}}
\newcommand{\ba}{\begin{eqnarray}}
\newcommand{\ea}{\end{eqnarray}}
\newcommand{\ban}{\begin{eqnarray*}}
\newcommand{\ean}{\end{eqnarray*}}

\newcommand{\eq}[1]{(\ref{#1})}

\begin{document}
\title{Acceleration of particles by Janis-Newman-Winicour singularities}

\author{Mandar Patil\footnote{Electronic address:mandarp@tifr.res.in}
and Pankaj S. Joshi\footnote{ Electronic address:psj@tifr.res.in}}

\affiliation{Tata Institute of Fundamental Research,
Homi Bhabha Road, Mumbai 400005, India.
}

\begin{abstract}
We examine here the acceleration of particles and high energy
collisions in the the Janis-Newman-Winicour (JNW) spacetime, which
is an extension of the Schwarzschild geometry when a massless
scalar field is included. We show that while the center of mass
energy of collisions of particles near the event horizon of
a blackhole is not significantly larger than the rest mass of
the interacting particles, in an analogous situation, it could
be arbitrarily large in the JNW spacetime near the naked singularity.
The high energy collisions are seen to be generic in the presence
of a photon sphere in the JNW spacetime, whereas an extreme
fine-tuning of the angular momentum of the colliding particles
is required when the photon sphere is absent. The center of mass
energy of collision near the singularity grows slowly for small
and extremely large deviations from the Schwarzschild blackhole,
but for intermediate strengths of the scalar field it rises
moderately fast. As a possible and potentially interesting
application, we point out that the presence of such high energy
collisions may help the blackhole configurations to be
distinguished from a naked singularity.

\end{abstract}
\pacs{04.20.Dw, 04.70.-s, 04.70.Bw}

\maketitle
\section{Introduction}
Various terrestrial particles accelerators collide particles
at the center of mass energies upto TeV scale which is significantly
below the Planck scale that corresponds to the quantum gravity
regime. Clearly, the physics in a very large energy range
remains unexplored today. In this connection, an intriguing
possibility would be to make use of various naturally occurring
processes in the vicinity of various astrophysical compact
massive objects where gravity would be very strong.

The blackholes are expected to provide such an environment.
With this motivation it was investigated whether it is possible
to have ultrahigh energy collisions in the vicinity of the
event horizon of the blackholes. Various compact very massive
objects that occur in the universe, for instance at the center
of most of the galaxies, are assumed to be Kerr blackholes
which are characterized by their spin $a$ and mass $M$.
In the geometrical units (where $c=G=1$),
The event horizon of such a blackhole is located at
$r=b(a)=M\left(1+\sqrt{1-a^2}\right)$ in the Boyer-Lindquist
coordinates. The center of mass energy of collision between two
particles interacting near the horizon released from infinity at
rest was investigated. When $a=0$ the blackhole is a Schwarzschild
blackhole and the maximum center of mass energy that can be
achieved in the collision at $r=b(a=0)=2M$ was shown to be finite
\cite{Bauchev}.
The effect of adding rotation to the blackhole was investigated in
\cite{BSW}.
It was shown that nothing interesting happens for small values
of spin. However, interestingly for large spins close to the
extremality $a=1$, it can be shown that the center of mass energy
of collision at $r=b(a=1)=M$ is arbitrarily large. Various aspects
of this process in Kerr and other blackhole geometries were investigated in
\cite{BSW2},\cite{Berti}.
We extended this process to super-spinning objects with spin
larger than unity containing a naked singularity
\cite{Patil},\cite{Patil2}.

Here we take a different approach to crank up the center
of mass energy of collision. Instead of spinning up a blackhole
we "charge it up" with a static massless scalar field. The
two parameter family solution of Einstein equations in this case
is given by
\cite{JNW}.
One parameter is mass $M$ and the second parameter is $q$,
the scalar charge which is a measure of strength of the scalar
field as we discuss in the next section. We investigate the
collision at $r=b(q)=2\sqrt{M^2+q^2}$. When $q=0$, the JNW metric
reduces to Schwarzschild metric and the center of mass energy
of collision at $r=b(q)=2M$ is small. We investigate here the
effect of invoking a scalar field on the center of mass energy
of collision at $r=b(q)=2\sqrt{M^2+q^2}$. We show that there
is a divergence of center of mass energy of collision. However,
the divergence is very slow for small and large values of $q$
and it is moderately fast for intermediate values of $q$
as $r=b$ is approached.

The JNW spacetime
\cite{JNW},
is an extension of the Schwarzschild geometry when
a massless scalar field is included instead of being empty.
Naively speaking, the blackhole event horizon then deforms
into a naked singularity in the JNW spacetime, when any
smallest non-zero value $q$ of the scalar field is included.
This metric was studied from the perspective of gravitational
lensing near a naked singularity
\cite{Vlens}.
If naked singularities which are hypothetical astrophysical
objects occur in nature, an important problem would be
how to distinguish them from the blackholes. Both these
entities which occur generically in gravitational collapse
of massive matter clouds within the framework of the
Einstein gravity
\cite{collapse},
present ultra-strong-gravity regimes which would
be clearly of much interest to study and explore the
physical effects of very strong gravity fields.

The JNW solution which always contains a naked
singularity has been studied extensively from the
perspective of gravitational lensing and distinguishing it
from blackholes
\cite{Vlens}.
It was shown that for certain range of parameters, in
the presence of a photon sphere, the gravitational lensing will
be qualitatively similar to that by the Schwarzschild blackhole
and it will not be possible to observationally distinguish
the blackhole from a naked singularity configuration. But
for the remaining range of the parameters in the absence of
the photon sphere, the observational signatures of the two in
terms of lensing will be quite different, allowing us to
differentiate between the two.

We show here that it would be possible to have ultrahigh
energy collisions generically in the presence of the photon sphere
in the case of a JNW naked singularity, whereas in case of the
Schwarzschild blackhole such collisions do not take place.
Conversion of dark matter and ordinary matter particles like
protons into particles with smaller mass, such as electrons,
would be efficient at large center of mass energy of collisions.
The flux of outgoing particles produced in collisions thus
is expected to be larger in the JNW spacetime as compared to
the Schwarzschild blackhole. We argue that this could lead
to the observational signature of the JNW naked singularities
in the presence of the photon sphere, as distinct from
the blackholes.

\section{The Schwarzschild and JNW metrics}
Schwarzschild metric is the unique solution to Einstein
equations obtained with the assumptions of spherical symmetry,
asymptotic flatness and absence of any source of the
energy-momentum, {\it i.e.} a vacuum geometry, and the solution
turns out to be static.  It is one of the simplest and most
widely studied exact solutions in the general theory of relativity.
The Schwarzschild metric in the coordinates adapted to spherical
symmetry and staticity $\left(t,r,\theta,\phi\right)$ is given
by the following expression,
\begin{equation}
ds^2=-\left(1-\frac{b}{r}\right)dt^2+\frac{1}{\left(1-\frac{b}{r}
\right)}dr^2+r^2d\Omega^2
\label{sch}
\end{equation}
The spacetime has a strong curvature singularity at
$r=0$ which is covered by an event horizon located at $r=b$
and thus it represents a blackhole. It is characterized by
the parameter $M=\frac{b}{2}$ which can be interpreted
as the mass which is the source of the gravitational field.

In this paper we investigate the unique solution to
the Einstein equations obtained with the assumption of 
spherical symmetry, asymptotic flatness
with the static massless scalar field as a source of energy-momentum
This solution was obtained by Janis, Newmann and Winicor
\cite{JNW}
and independently by Wyman
\cite{Wyman},
and these were later shown to be identical
\cite{VJJ}.

The JNW metric is given by
\begin{equation}
ds^2=-\left(1-\frac{b}{r}\right)^{\nu}dt^2+\frac{1}
{\left(1-\frac{b}{r}\right)^{\nu}}dr^2+r^2 \left(1-\frac{b}{r}\right)
^{1-\nu}d\Omega^2
\label{JNW}
\end{equation}
whereas the scalar field here is given by
\begin{equation}
\phi=\frac{q}{b\sqrt{4\pi}} \ln\left(1-\frac{b}{r}\right)
\label{JNWc}
\end{equation}
The solution contains two parameters
\begin{eqnarray}
\nu=\frac{2M}{b} \\
\nonumber
b=2\sqrt{M^2+q^2}
\end{eqnarray}
Here $M$ and $q$ stand for the ADM mass and the `scalar charge'
respectively. When the scalar field is zero the solution should
reduce to the Schwarzschild solution. This indeed happens and can
be seen clearly by setting $q=0$, {\it i.e.} $\nu=1$
in which case \eq{JNW} reduces to \eq{sch}. The JNW solution differs
from the Schwarzschild solution for non-zero values of the
scalar charge. As we increase $q$, $\nu$ goes on decreasing
and tends to zero for arbitrarily large values of the charge.
The energy-momentum tensor of the scalar field is given
here by
\begin{eqnarray}
T^{\mu}_{\nu}=Diag\left[-\rho(r),p_{1}(r),p_{2}(r),p_{2}(r)\right] \\
\nonumber
\rho(r)=p_{1}(r)=-p_{2}(r)=\frac{b^2 (1-\nu^2)\left(1-\frac{b}{r}
\right)^{\nu}}{4r^2\left(r-b\right)^2}
\end{eqnarray}

It is clear from the expression above that the JNW spacetime
has a singularity at $r=b$ when $\nu<1$. Thus the range of the
radial coordinate is given by $b<r<\infty$. The singularity
can be shown to be globally naked
\cite{VJJ},\cite{VB}.
The weak energy conditions are satisfied.

Thus to summarize, the JNW metric represents the two
parameter family of solutions to the Einstein equations that
is spherically symmetric and static. The two parameters
$b,\nu$, or alternatively $M,q$ represent the mass and scalar
charge. The Schwarzschild blackhole is the limiting case
of JNW spacetime when $q=0$, {\it  i.e.} when $\nu=1$.
When $q \neq 0$ {\it i.e.} $0<\nu<1$, then $r=b$ is the naked
singularity, whereas in the limiting case of the Schwarzschild
blackhole $q=0$ ($\nu=1$), it represents the event horizon
of the blackhole.
Thus one might be tempted to dramatically
assert that, introduction of the smallest possible static
scalar field in the Schwarzschild blackhole could convert
it into a JNW spacetime containing a naked singularity,
and the event horizon at $r=b$ is then deformed into
a naked singularity.
However, while considering the statement above
we must note that the JNW spacetime is not to be thought 
of as a perturbation of
the Schwarzschild spacetime in the usual sense, {\it i.e.} 
the variation of initial data for
Schwarzschild spacetime on any initial surface 
does not give rise to the JNW metric.
It can be thought of as a variation of the Schwarzschild metric 
in the sense that we describe below.

Schwarzschild solution is obtained by solving the system of Einstein equations
with the assumptions of spherical symmetry, asymptotic flatness and the absence of
matter, {\it i.e.} a vacuum geometry.
The solutions to Einstein equations obtained by relaxing one or more assumptions
above are different and have different physical interpretations.
If one sticks to the assumptions of spherical symmetry and asymptotic flatness and
introduces matter fields, one would obtain a wide variety of solutions.
For instance, we arrive at Reissner-Nordstar\"{o}m geometry if 
one invokes the static electromagnetic field as a source of the energy-momentum.
If we invoke the static massless scalar field we get JNW solution.  
In both these cases the departure
from the Schwarzschild geometry is characterized by one parameter. 
In the case of electromagnetic field the parameter is an electric
charge. By analogy in the case of scalar field, the additional
parameter is referred
to as the scalar charge. If the additional charge parameter is set 
to zero we recover the Schwarzschild geometry.
However, there is a significant difference in the two cases.
If the electric charge is small, the Reissner-Nordstr\"{o}m geometry 
corresponds to a blackhole as for the asymptotic observers are 
concerned, and there is an event horizon present in the spacetime. 
However, the spacetime contains a locally naked singularity, visible 
in a region inside the Cauchy horizon which is the inner event
horizon, but invisible to the observers outside the horizon.
In contrast, in the case of JNW solutions the geometry corresponds 
to that of a globally naked singularity, rather than
a blackhole, even for a smallest possible scalar charge, and
there is no event horizon in the spacetime.
This is the reason we investigate JNW solution in this paper 
as a representative of the variations of the Schwarzschild geometry
that yield naked singularities rather than blackholes.

Looking from the sufficiently faraway region from the center, 
{\it i.e.} by observing the phenomenon that takes place in the weak
field 
regime, like the motion of stars or planets, one can infer only the 
mass parameter for the system. Therefore the JNW solution essentially looks like
Schwarzschild solution from such a perspective and one may not be able 
to distinguish between the two.
So the compact massive dark object, that we might usually believe to
be a blackhole, could turn out to be a naked singularity if the scalar 
field is present instead of it being vacuum.
In order to tell one from the other, it would be necessary to 
investigate various physical
phenomena that take place in the vicinity of this object, that is 
in the strong gravity regime, like the accretion of matter onto the 
object or gravitational lensing. This would shed light on the
existence of the charge 
parameter and therefore the deviation of metric from being 
Schwarzschild. From this perspective,
gravitational lensing by JNW singularities was studied in 
\cite{Vlens}. We study and present the analysis on the circular 
geodesics and accretion disks in the JNW geometry in a separate 
companion paper \cite{Patil3}.

In this paper we investigate the effect of the inclusion
of the massless static scalar field on the acceleration of particles
and their collisions occurring at $r=b$.

\section{Particle acceleration near a Schwarzschild blackhole}
In this section we describe the particle acceleration
and collisions near the event horizon in the Schwarzschild
spacetime. The event horizon of a blackhole is chosen to be
a location for collision because it is a surface with infinite
blueshift for the particles falling in from infinity into
the blackhole. Thus naively one might think that when
two such highly blueshifted particles collide near the horizon,
the center of mass energy of collisions will be arbitrarily
large. But actually that is not the case as we show below.

\subsection{Geodesic motion}
We first describe motion of the massive particles
following timelike geodesics in the Schwarzschild spacetime.
Consider a particle of mass $m$.
Let $U^{\mu}=\left(U^t,U^r,U^{\theta},U^{\phi}\right)$ be the
four-velocity of the particle. The motion of the particle will
be restricted to a plane because of the spherical symmetry of
a spacetime. This plane can be chosen to be the equatorial
plane $\theta=\frac{\pi}{2}$, exploiting the gauge freedom.
Thus $U^{\theta}=0$.  The spherical symmetry and static nature
of the Schwarzschild spacetime implies that it admits
Killing vectors $\partial_{\phi},\partial_{t}$.
Using these Killing vectors one can show that the quantities
$E=-\partial_{t}\cdot U$ and $L=-\partial_{\phi}\cdot U$ turn out
to be the constants of motion along the geodesics.
Here $E,L$ are interpreted as the conserved energy and
angular momentum per unit mass of the particle respectively.
The quantities $U^t,U^{\phi}$ can be written in terms
of $E,L$ in the following way,
\begin{eqnarray}
\nonumber
U^{t}=\frac{E}{\left(1-\frac{b}{r}\right)} \\
U^{\phi}=\frac{L}{r^2}
\label{Utfs}
\end{eqnarray}
Using the normalization condition $U.U=-1$, $U^{r}$
can be written in the following way,
\begin{equation}
U^r=\pm \sqrt{E^2-\left(1-\frac{b}{r}\right)\left(1+\frac{L^2}{r^2}\right)}
\label{Urs}
\end{equation}
Here $\pm$ correspond to the motion in radially outward
and inward directions respectively.
This equation can also be cast in the following form
\begin{eqnarray}
(U^r)^2+V_{eff}(r)=E^2 \\
\nonumber
V_{eff}(r)=\left(1-\frac{b}{r}\right)\left(1+\frac{L^2}{r^2}\right)
\label{UrsV}
\end{eqnarray}
where $V_{eff}(r)$ can be thought of as an effective
potential for the motion in the radial direction.

We are interested here in a situation where particles
that are nonrelativistic at infinity fall inwards under the
force of gravity and interact with each other with large center
of mass energy of collision. The particles that are non-relativistic
at infinity can be thought to be at rest for all
practical purposes. Then $U^r\rightarrow 0$ as $r \rightarrow \infty$
implies from \eq{Urs},\eq{UrsV} that the conserved energy of the
particles must be $E=1$. We impose this
condition to ensure that if particles undergo ultrarelativistic
collision, the large center of mass energy of collision
can be attributed purely as an effect of
infall under the gravity.

As we have mentioned before, intuitively the location
of the collision must be close to the event horizon as it
is a surface with infinite blueshift for the ingoing particles.
Thus we must ensure that the particles that fall from rest
at infinity manage to reach
the event horizon and that they do not turn back well before the
same is reached. This can be done in a following way.

Consider a particle following geodesic motion that
turns back at $r$. This implies that $U^{r}=0$ or $V_{eff}(r)=E^2=1$.
From \eq{Urs},\eq{UrsV}, we get
the expression of the angular momentum $L(r)$, and
such a particle is required to have
\begin{eqnarray}
L^2(r)=\frac{r^2b}{r-b} \\
\nonumber
L(r)=\pm \sqrt{\frac{r^2b}{r-b}}
\end{eqnarray}

\begin{figure}
\begin{center}
\includegraphics[width=0.5\textwidth]{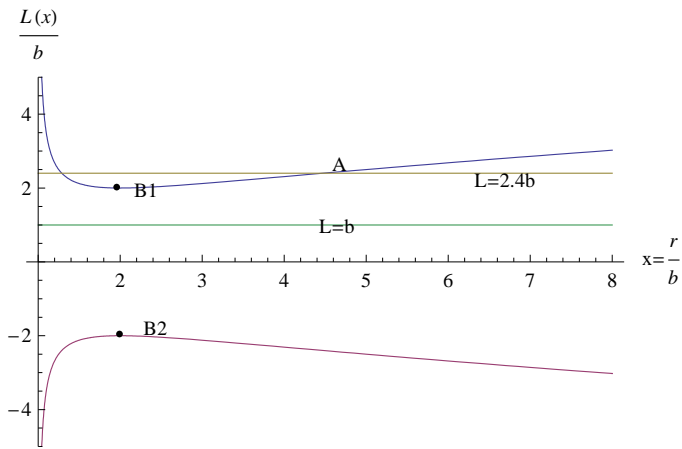}
\caption{\label{fg1}
The angular momentum required for a particle to turn back
at radius $r$, namely $\frac{L}{b}$ is plotted against
$x=\frac{r}{b}$. There are two branches of the graph,
namely the positive and negative branches, which admit minimum
and maximum at $B1, B2$ respectively. The extrema occur
at $r=2b$ and the angular momenta take extremal values $\pm2b$.
An infalling particle with angular momentum $L=b$ never turns
back and reaches the event horizon, whereas the ingoing particle
with an angular momentum $L=2.4b$ which is outside this
range gets reflected back at point $A$ with the radial
coordinate $r>2b$.
}
\end{center}
\end{figure}

The function $L(r)$ is plotted in Fig1. There are two branches
of the graph corresponding to the positive and negative values of
the angular momenta.
The positive branch admits minimum, whereas the negative branch
admits maximum at $r=2b$. The extremal values are $-2b,2b$.
If the angular momentum of the particle is in the range
$L\in (-2b,2b)$, it never turns back and reaches the horizon.
But if the angular momentum is outside this range {\it i.e.}
when either $L>2$ or $L<-2$, an ingoing particle would turn back
at the radial value $r>2$. A particle with angular momentum
$L=\pm 2b$ would asymptotically approach $r=2b$. This follows
from the fact that for such a particle
$V_{eff}(r=2)=V_{eff}'(r=2)=0$, where
the prime denotes the derivative
with respect to $r$. Thus for particles infalling from
infinity at rest to participate in the collisions at the event
horizon of the Schwarzschild blackhole, their
angular momentum must be in the range mentioned
as above.

\subsection{Collisions and Center of Mass Energy}
The center of mass energy of collision \cite{Bauchev},\cite{BSW} between two identical
particles each with mass $m$ and with velocities $U_1,U_2$ is given by
\begin{equation}
E_{cm}^2=2m^2\left(1-g_{\mu \nu}U_1^{\mu}U^{\nu}_2\right)
\label{ecm}
\end{equation}
The calculation of center of mass energy of collision
essentially involves the computation of the inner product of
the velocities of two particles.

We consider two particles each with the conserved energy
$E=1$ and angular momenta $L_1,L_2$ satisfying the condition
$-2b<L_1,L_2<2b$. These particles are released
from infinity at rest and manage to reach the event horizon
to participate in the collision.

The center of mass energy of collision between these two
particles at any given value of $r$ can be computed using
\eq{Utfs},\eq{Urs},\eq{ecm} and is given by,
\begin{eqnarray}
\nonumber
\frac{E_{cm}^2}{2m^2}=1+\frac{1}{\left(1-\frac{b}{r}\right)}-
\frac{1}{\left(1-\frac{b}{r}\right)}\sqrt{1-\left(1-\frac{b}{r}
\right)\left(1+\frac{L_1^2}{r^2}\right)}\\
\sqrt{1-\left(1-\frac{b}{r}\right)\left(1+\frac{L_2^2}{r^2}\right)}
-\frac{L_1L_2}{r^2}
\end{eqnarray}
We consider a collision close to the event horizon $r=b$.
The second term under each of the square root in the expression
above is much smaller than the first term.
Taylor expanding the square root and keeping terms upto
the first order we obtain the center of mass energy of collision
close to the horizon as given by,
\begin{equation}
E_{cm}^2=2m^2\left(2+\frac{\left(L_1-L_2\right)^2}{2r^2}\right)
\end{equation}

The center of mass energy of collision between the two
particles is maximum when angular momenta are opposite in sign
and take extreme values, {\it i.e.} $L_1=2b,L_2=-2b$.
\begin{equation}
E_{cm,max}^2=20m^2
\end{equation}
Thus the center of mass energy of collision between two
particles at horizon $r=b$ for a Schwarzschild blackhole is not
significantly large as compared
to their rest mass. This result can be intitutively
understood in the following way.
Although the two particles are individually highly
blue-shifted as they reach the horizon, they move almost parallel
to each other. Their relative velocity is small.
Therefore the center of mass energy of collision is small.

There are a number of ways to crank up and raise
the center of mass
energy of collision. For instance one could spin up or charge up
the blackhole. Nothing interesting happens really even in the case of
the Kerr or Reisnner-Nordstrom blackholes unless one reaches
the extremal limit.
However, the center of mass energy of collision between the
particles is shown to be divergent near the event horizon
of extremal or near extremal blackholes
\cite{BSW}\cite{BSW2},
provided certain fine-tuning assumptions are obeyed.
We also extended the particle acceleration mechanism to near
extremal naked singularities
\cite{Patil},\cite{Patil2}.

We note that one of the main drawbacks of the particle acceleration
process by near extremal blackhole has been that the extreme
finetuning of the angular momentum of one the particles
was required for the divergence of the center of mass energy
of collision.

In this paper we suggest a different way to crank up the
center of mass energy of collision by "charging up" the Schwarzschild
blackhole with a static spherically symmetric scalar field.
By doing that we obtain the JNW solution. We study the particle
acceleration in the JNW metric in the next section.

\section{Particle acceleration in JNW spacetimes}
In the previous section we investigated the collision of
the particles near the event horizon of the blackhole $r=b$,
and showed that the center of mass energy of collision between
the particles was not significantly larger compared to the
rest mass. In this section we study the collision between the
particles at $r=b$ in the JNW spacetime obtained from
the Schwarzschild geometry by inclusion of the massless scalar
field. We show that the center of mass energy of collision
in JNW spacetime can be arbitrarily large. Thus inclusion of
the scalar field changes the phenomenon of particle acceleration
significantly. We also investigate here the genericity of
the ultrahigh collision process in terms of the allowed values
of the geodesic parameters associated with the particles
participating in the ultrahigh energy collisions.

\subsection{Geodesic motion in JNW spacetime}
We describe here the motion of the massive particles following
the timelike geodesics in the JNW spacetime. Consider a particle of
mass $m$ and four velocity
$U^{\mu}=\left(U^t,U^r,U^{\theta},U^{\phi}\right)$. As in the
case of Schwarzschild blackhole, this particle moves in a plane
which can be taken to be
an equatorial plane, and thus $U^{\theta}=0$. Its motion can be
described in terms of the constants of motion $E=-\partial_{t}.U$
and $L=\partial_{\phi}.U$, namely the
conserved energy and angular momentum per unit mass
of the particle. Using the constants of motion and \eq{JNW},
$U^t,U^{\phi}$ can be written as,
\begin{eqnarray}
\nonumber
U^{t}=\frac{E}{\left(1-\frac{b}{r}\right)^{\nu}} \\
U^{\phi}=\frac{L}{r^2\left(1-\frac{b}{r}\right)^{1-\nu}}
\label{Utfj}
\end{eqnarray}

Using the normalization condition for the velocity,
namely $U.U=-1$, $U^{r}$ can be written as
\begin{equation}
U^r=\pm \sqrt{E^2-\left(1-\frac{b}{r}\right)^{\nu}
\left(1+\frac{L^2}{r^2\left(1-\frac{b}{r}\right)^{1-\nu}}\right)}
\label{Urj}
\end{equation}
Here $\pm$ correspond to the motion in radially outward
and inward directions respectively.
This equation can also be written in the following form
\begin{eqnarray}
(U^r)^2+V_{eff}(r)=E^2 \\
\nonumber
V_{eff}(r)=\left(1-\frac{b}{r}\right)^{\nu}\left
(1+\frac{L^2}{r^2\left(1-\frac{b}{r}\right)^{1-\nu}}\right)
\label{UrjV}
\end{eqnarray}
where $V_{eff}(r)$ can be thought of as an effective potential
for the motion in the radial direction.

We set $E=1$ since we consider particles that are released
from infinity at rest and which fall freely under the gravity.
Here "infinity" corresponds to $r\rightarrow \infty$ subject to $\nu \in (0,1)$. Since we
are interested in the collisions of the particles at $r=b$, we
must ensure that the particles are not reflected back
before they reach there.

Consider a particle that turns back at a given radial
coordinate $r$. We must have $U^{r}(r)=0$ or $V_{eff}(r)=1$.
It follows from \eq{Urj},\eq{UrjV} that if the particle were
to turn back from $r$ it must have an angular momentum
which is given by,
\begin{eqnarray}
\nonumber
L^2=\left[\left(1-\frac{b}{r}\right)^{1-2\nu}-
\left(1-\frac{b}{r}\right)^{1-\nu}\right]r^2 \\
L=\pm\sqrt{\left[\left(1-\frac{b}{r}\right)^{1-2\nu}-
\left(1-\frac{b}{r}\right)^{1-\nu}\right]}r
\label{am1}
\end{eqnarray}
There are two branches of the angular momentum,
namely the positive branch and negative branch.

The parameter $\nu$ ranges from the values $0$ and $1$.
For all values of $\nu$ the power of the second term in the
parenthesis $\left(1-\nu\right)$
is nonnegative whereas the power of the first term
$\left(1-2\nu\right)$ can either be positive, negative or
zero depending on whether $\nu<\frac{1}{2}$,$\nu>\frac{1}{2}$
or $\nu=\frac{1}{2}$. Thus as we approach $r=b$ the first term
can either go to zero, blow up or remain finite.
We analyze these three cases separately here. Interestingly,
as we discuss in the next section, $\nu=\frac{1}{2}$ also marks
the disappearance of the photon sphere in the spacetime.
For $\nu<\frac{1}{2}$ the photon sphere is absent, whereas
for $\nu>\frac{1}{2}$, photon sphere is present in
the spacetime.

\subsubsection{Geodesics in JNW spacetime with $\nu<\frac{1}{2}$}
We first discuss the case when $\nu<\frac{1}{2}$. We focus
on the positive branch of the angular momentum. The discussion
for the negative branch will be identical
to that of the positive branch due to the mirror symmetry.
For large values of $r$, the angular momentum required for
the particle to turn back has a following behavior,
\begin{equation}
L\approx \sqrt{\nu r b}
\end{equation}
On the other hand, for values of $r\rightarrow b$ we have,
\begin{equation}
L\rightarrow 0
\end{equation}
Thus for a particle to reach near $r=b$ and participate
in the collision it must have an angular momentum tending
to zero. Thus we need a fine-tuning of the angular
momentum of both the particles to a vanishingly small value
for them to reach arbitrarily close to $r=b$ and
participate in the collision.

\subsubsection{Geodesics in JNW spacetime with $\nu=\frac{1}{2}$}
For the case $\nu=\frac{1}{2}$, angular momentum required
for the particle to turn back at $r$ is given by
\begin{equation}
L(r)=r\sqrt{1-\left(1-\frac{b}{r}\right)^{\frac{1}{2}}}
\end{equation}
We are dealing with the positive branch here and
this is plotted in Fig2.

For large values of $r$ the angular momentum goes as
\begin{equation}
L(r)\approx \sqrt{\frac{br}{2}}
\end{equation}
whereas at $r=b$ it attains a constant value
\begin{equation}
L(r)\approx b
\end{equation}
It admits a local minimum at $r=1.125$ and the minimum
value of the angular momentum is given by
\begin{equation}
L_{min}=0.844 b
\end{equation}
By mirror symmetry the negative branch will admit a maximum
at the same value of the radial coordinate and the value of the
angular momentum will be the negative of the minimum
for the positive branch.

\begin{figure}
\begin{center}
\includegraphics[width=0.5\textwidth]{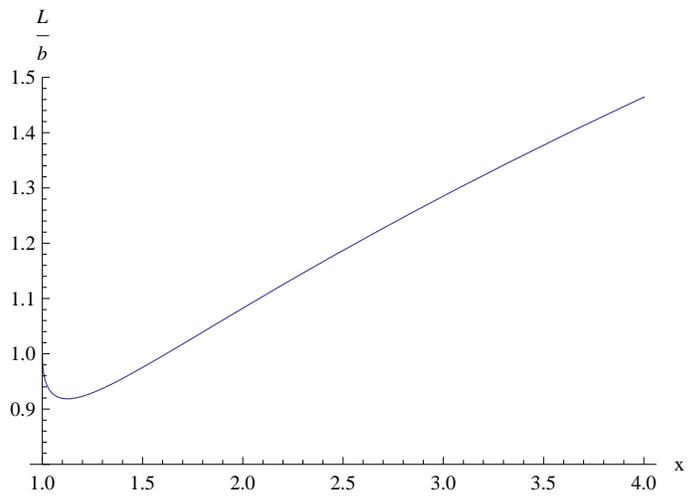}
\caption{\label{fg2}
The positive branch of angular momentum required for a
particle to turn back at $r$, namely $\frac{L}{b}$ is plotted
against $x=\frac{r}{b}$. At $r=b$,
the angular momentum takes a finite value. It admits a minimum
at $r=0.125b$ and the minimum value is $L_{min}=0.84b$.
Particles traveling inwards would reach $r=b$ if their
angular momentum is in the range $L \in (-L_{min},L_{min})$.
Otherwise they would turn back.
}
\end{center}
\end{figure}

Thus the ingoing particles from infinity with angular
momentum in the range
\begin{equation}
-0.844b<L<0.844b
\label{Lrang}
\end{equation}
will be able to reach $r=b$ and participate in the
collisions. Whereas particles with the angular momentum
outside this range will get deflected away
from $r>1.125b$.

\subsubsection{Geodesics in JNW spacetime with $\nu>\frac{1}{2}$}
Finally we consider the third case where $1>\nu>1/2$.
We analyze the positive branch of the angular momentum for
the particle to turn back. For large values of $r$, as earlier
angular momentum has a following behavior
\begin{equation}
L\approx \sqrt{\nu r b}
\end{equation}
For small values of $r\rightarrow b$ we get
\begin{equation}
L\rightarrow \infty
\end{equation}
It admits a minimum for an intermediate value of $r$.
It is difficult to analyze the minimum analytically.
We plot in Fig3 the angular momentum against $r$ on x-axis
and $\nu$ on y-axis.

\begin{figure}
\begin{center}
\includegraphics[width=0.5\textwidth]{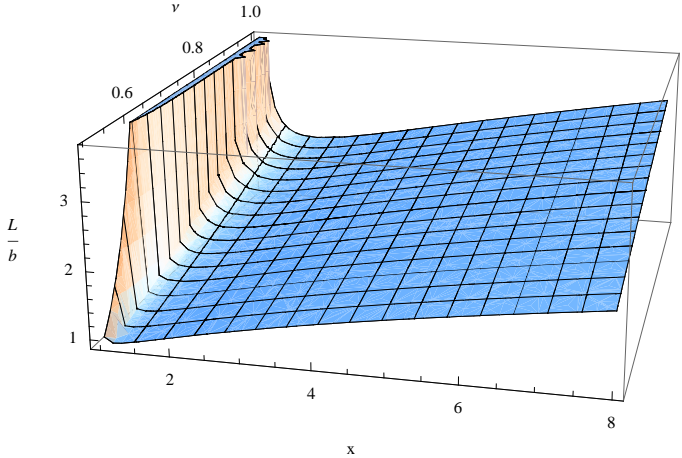}
\caption{\label{fg3}
The positive branch of angular momentum required for a
particle to turn back at $r$, namely $\frac{L}{b}$ is plotted
against on x-axis $x=\frac{r}{b}$ and on y-axis
$\nu$. The range of $\nu$ is from $0.5$ to $1$. The angular
momentum blows up at $r=b$. The value of radial coordinate $r$
where minimum of the angular momentum is attained and the
minimum value of the angular momentum goes on
increasing with increasing $\nu$.
}
\end{center}
\end{figure}

It can be clearly seen that the radial coordinate at which
the minimum is attained, as well as the minimum value of the angular
momentum, increases as we increase $\nu$ from $0.5$ to $1$.
Thus it follows that if the angular momentum of the ingoing
particle is in the range
\begin{equation}
-0.844b<L<0.844b
\label{Lrang2}
\end{equation}
it would definitely reach $r=b$ to participate in the
collision for various values of $\nu$.

\subsection{Collisions and Center of Mass Energy}
We now describe the acceleration collision of particles in
JNW spacetime. We show that unlike the Schwarzschild blackhole
case it is possible to have high energy collsions
at $r=b$ when the scalar field is nonzero. We also examine the
genericity of the process in terms of the allowed range of
the angular momentum of the colliding particles
participating in the high energy collisions.

\subsubsection{Case $0<\nu <\frac{1}{2}$}
We first consider the case where $\nu <\frac{1}{2}$. Two
identical particles of mass $m$ are released from infinity at rest.
If they were to reach and interact in the vicinity of $r=b$
they must have a vanishingly small angular momentum. We
assume that the collision
taken place at $r$ close to $r=b$ and two particles have angular
momenta given by \eq{am1} $L(r)\rightarrow 0$ that would make
their component of radial velocity $U^{r}$ to be identically zero.
We assume that one of the particles has a positive and the other
has a negative angular momentum. This is because if they have the
same angular momentum, they have identical velocities, their
relative velocity is zero and thus the center of mass energy
of collision will be $2m$ .

The center of mass energy of collision between the
particles,
using \eq{ecm},\eq{Utfj},\eq{Urj} is given by,
\begin{eqnarray}
E_{cm}^2=2m^2\left(1-g_{\mu\nu}U_{1}^{\mu}U_{2}^{\nu}\right)
\end{eqnarray}
\begin{equation}
\nonumber
=2m^2\left(1+\frac{1}{\left(1-\frac{b}{r}\right)^{\nu}}+
\frac{L(r)^2}{\left(1-\frac{b}{r}\right)^{1-\nu}r^2}\right)
\end{equation}
Using \eq{am1} we get
\begin{eqnarray}
E_{cm}^2=\frac{4m^2}{\left(1-\frac{b}{r}\right)^{\nu}} \\
\nonumber
\lim_{r \to b}E_{cm}^2 \rightarrow \infty
\end{eqnarray}
It follows that the center of mass energy of collision
blows up as we approach $r=b$.

Fig4 shows the variation of center of mass energy of collision
as a function of radial coordinate and $\nu$.
For small values of $\nu$ the divergence is slow and one has
to go very close to $r=b$ to obtain sufficiently large center
of mass energies.
But for $\nu \approx 0.5$ we get large center of mass energies
at moderate distance from $r=b$.

One limitation of the particle acceleration process for this
range of parameter values for $\nu$ is that the angular momentum must be
arbitrarily close to being zero.
Thus the fine-tuning of angular momentum is necessary.
Values of parameters in this range correspond to large
deviation from the Schwarzschild geometry. It turns out that
it is possible to have high energy collisions, however,
this process is very much fine-tuned.

\subsubsection{Case $\frac{1}{2}\le \nu <1$}
We now study the particle collisions for the remaining
values of the parameter $\nu$, namely $\frac{1}{2}\le \nu <1$.
We present a combined analysis of cases $\nu=\frac{1}{2}$
and $\frac{1}{2}<\nu<1$ which we had analyzed previously because
there was a qualitative difference in the behavior of the
angular momentum function $L(r)$ near $r=b$. However, while
calculating the center of mass energy of collision they
can be analyzed together.

We again consider two particles released from infinity at
rest with the angular momenta in the range given by\eq{Lrang}
so that they reach $r=b$ and participate
in the collision. The center of mass energy of collision
between these two particles \eq{ecm},\eq{Utfj},\eq{Urj} is given by,
\begin{eqnarray}
E_{cm}^2=2m^2\left(1-g_{\mu\nu}U_{1}^{\mu}U_{2}^{\nu}\right)
\end{eqnarray}
\begin{eqnarray}
\frac{E_{cm}^2}{2m^2}=1+\frac{1}{\left(1-\frac{b}{r}\right)^{\nu}}-
\frac{L_1L_2}{r^2\left(1-\frac{b}{r}\right)^{1-\nu}}-\frac{1}
{\left(1-\frac{b}{r}\right)^{\nu}}\\
\nonumber
\sqrt{1-\left(1-\frac{b}{r}\right)^{\nu}\left(1+\frac{L_1^2}
{r^2\left(1-\frac{b}{r}\right)^{1-\nu}}\right)}\\
\nonumber
\sqrt{1-\left(1-\frac{b}{r}\right)^{\nu}\left(1+\frac
{L_1^2}{r^2\left(1-\frac{b}{r}\right)^{1-\nu}}\right)}
\end{eqnarray}

The second term under the square root close to $r=b$ is
negligible as compared to the first term. This follows from
the fact that $\nu>\frac{1}{2}$. Thus Taylor expanding the square
root and neglecting the terms beyond first order we get,
\begin{eqnarray}
E_{cm}^2=2m^2\left(2+\frac{1}{2}\frac{\left(L_1-L_2\right)^2}
{2\left(1-\frac{b}{r}\right)^{1-\nu}r^2}-\frac{L_1^2L_2^2 \left(1-
\frac{b}{r}\right)^{3\nu-2}}{r^4}\right) \\
\nonumber
\lim_{r \to b} E_{cm}^2 \rightarrow \infty
\end{eqnarray}
The third term in the above is important only when
$\nu \in (\frac{1}{2},\frac{2}{3})$, and it can be neglected for
the remaining values of the parameter $\nu$. Even in this range,
the second term dominates the first term since
$\left(1-\nu\right)>\left(2-3\nu\right)$.

The center of mass energy diverges as $r=b$ is approached.
The divergence is very slow for $\nu\approx 1$, whereas it is
moderately fast for $\nu \approx \frac{1}{2}$.

The process of particle acceleration and high energy collision
is generic for this range of values of parameters since the angular
momenta of the particles participating in the collision lie in
a finite range, unlike those taking a single
fine-tuned value.

\section{Observationally distinguishing the Schwarzschild blackhole
from JNW naked singularity}
Gravitational lensing has been suggested as a way of
distinguishing the blackholes from JNW naked singularities that
occur if the scalar field is present in the spacetime
\cite{Vlens}.
The gravitational lensing phenomenon, however, depends
crucially on whether or not a photon sphere is present in the
spacetime. In the presence of the photon sphere
gravitational lensing by the JNW spacetime turns out to be
qualitatively similar to that by the Schwarzschild blackhole.
However, in the absence of the photon sphere the gravitational
lensing in the JNW geometry is qualitatively different
from the blackhole case.

The photon sphere is present in JNW spacetime when
$\frac{1}{2}<\nu<1$, whereas it is absent when $0<\nu<\frac{1}{2}$.
Thus gravitational lensing as discussed in \cite{Vlens},will be a useful observational technique
to distinguish blackholes from JNW naked singularity when
$0<\nu<\frac{1}{2}$, and it will be essentially ineffective
when the scalar field strengths lie in the range
$\frac{1}{2}<\nu<1$.

However, as we have now shown in the previous section,
it is possible to have high energy particle collisions around
the JNW naked singularity for the range of values
$\frac{1}{2}<\nu<1$. On the other hand, the particle collisions
around the blackhole will be low center of mass energy collisions.

What this means is, a naked singularity can have different
possible manifestations in terms of the physical effects it may
generate. The gravitational lensing could be one such effect
which may help us differentiate the blackholes from a naked
singularity which is a hypothetical astrophysical object,
when it occurs in nature. But when that fails, for example, in
the case of presence of a photon sphere in the case of the
JNW solution, then there could be other physical effects
such as the acceleration and high energy collision of the
particles, which could give characteristic signatures to
differentiate the naked singularity from a blackhole.

The cross-sections for various particle physics processes
depend on the center of mass energy of collisions between the
particles. It increases typically with the center of mass energy.
Thus there would be a characteristic difference between the
collisions that occur around the blackhole and the JNW naked
singularities. Dark matter as well as ordinary matter particles
like protons would be falling in towards the central massive
object due to strong gravity, and hence must undergo collisions.
In the case of blackholes the collisions will be small center
of mass energy collisions, whereas in case of JNW naked singularities
the center of mass energy collisions will be very large.
Since the cross section for various particle physics processes
in general would be large at the large center of mass energy of
collisions, therefore the flux of the outgoing particles like
electrons, neutrinos and so on which have been produced in
the collisions could be expected to be quite large in the case
of JNW naked singularities as compared to their blackhole
counterparts.

Such a scheme can then be possibly used to observationally
distinguish blackholes and the JNW naked singularities with the
parameter range $\frac{1}{2}<\nu<1$, where the gravitational lensing
proves ineffective for the purpose of distinction of the two due
to the presence of the photon sphere.

\section{Concluding remarks}
In this paper we studied the effect of inclusion of a scalar
field onto the center of mass energy of collision between the
particles. The main purpose here was to show that it is possible
to have ultrahigh energy collisions in the JNW spacetime in the
presence of a static massless scalar field around naked singularities,
while the high energy collisions are absent in the Schwarzschild
blackhole case when the scalar field is zero.

We also speculated that the high energy collisions could allow
us to set up a scheme to observationally distinguish the Schwarzschild
blackhole from the JNW naked singularities which are surrounded by
the photon sphere, in the case when the gravitational lensing is
ineffective for such a purpose. It is beyond the scope of this paper
to rigorously demonstrate it here and we plan to report that
work elsewhere.

Also, in the future work we would like to study the process
of infall of dark matter particles and also ordinary particles like
protons onto the galactic central dark object, modeled as a
blackhole and as a JNW spacetime with a photon sphere surrounding
the naked singularity. A good understanding of the density of
dark matter and ordinary matter particles in the surroundings
of these objects will have to be obtained and studied. We intend
to study the collisions and investigate various particle physics
processes leading to the annihilation of dark and ordinary matter
particles into other particles like neutrinos and such other
by-products. Finally, one would like to calculate the flux of
the particles produced in the collisions at the large distances
for JNW naked singularities, and compare it with the
corresponding flux in the blackhole case.

\end{document}